\documentclass[11pt,a4paper,english,showpacs,superscriptaddress,prd,aps,
preprint]{revtex4}

\usepackage{amssymb}
\usepackage{amsmath}
\usepackage{graphicx}
\usepackage{babel}
\usepackage{hyperref}
\usepackage{color}

\begin{document}

\title{Supergauge theories in aether superspace}

\author{A.~C.~Lehum}
\email{andrelehum@ect.ufrn.br}
\affiliation{Instituto de F\'\i sica, Universidade de S\~ao Paulo\\
Caixa Postal 66318, 05315-970, S\~ao Paulo, SP, Brazil}
\affiliation{Escola de Ci\^encias e Tecnologia, Universidade Federal
  do Rio Grande  do Norte\\
Caixa Postal 1524, 59072-970, Natal, Rio Grande do Norte, Brazil}

\author{J. R. Nascimento}
\email{jroberto,petrov@fisica.ufpb.br}
\affiliation{Departamento de F\'{\i}sica, Universidade Federal da 
Para\'{\i}ba\\
 Caixa Postal 5008, 58051-970, Jo\~ao Pessoa, Para\'{\i}ba, Brazil}

\author{A. Yu. Petrov}
\email{jroberto,petrov@fisica.ufpb.br}
\affiliation{Departamento de F\'{\i}sica, Universidade Federal da 
Para\'{\i}ba\\
 Caixa Postal 5008, 58051-970, Jo\~ao Pessoa, Para\'{\i}ba, Brazil}

\author{A. J. da Silva}
\email{ajsilva@fma.if.usp.br}
\affiliation{Instituto de F\'\i sica, Universidade de S\~ao Paulo\\
Caixa Postal 66318, 05315-970, S\~ao Paulo, SP, Brazil}

\begin{abstract}
Within the superfield method we extend the formulation of the
Lorentz-breaking aether superspace for supergauge theories, both in
three- and in four-dimensional  cases.  


\end{abstract}

\maketitle

\section{Introduction}\label{intro}

The idea of the Lorentz symmetry breaking has been intensively
discussed during the last years (for a review, see
f.e. \cite{Bietenholz}). The interest in this line of studies is
motivated by the fact that the presence of the
Lorentz-breaking additive  modifications of field theory models,
essentially enriches their structure \cite{Kostel}. At the same time,
since supersymmetry is treated as a fundamental physical symmetry, a
natural question is -- whether the Lorentz-breaking field theory can
be supersymmetric. A systematic methodology used to address this
problem is based on the Kostelecky-Berger construction \cite{KostBer}
involving the deformation of the supersymmetry algebra, which, in
principle, can be applied to different kinds of superfield theories,
formulated in different space-time dimensions, and allows for the arising
of the CPT-even Lorentz-breaking terms on the component level. Other
possible solution for this problem involve  extra superfield(s) whose
component(s) depend on the Lorentz-breaking parameters \cite{Hel}
(which can allow for the arising of the CPT-odd Lorentz-breaking terms
on the component level), or a straightforward addition of
Lorentz-breaking, superfield-dependent terms like
$k^{ab}\partial_a\Phi\partial_b\bar{\Phi}$ (where $\Phi$, $\bar{\Phi}$
are the superfields, and the $k^{ab}$ is a constant tensor). To the
best of our knowledge, however, the last method has not been
systematically used yet, and it is clear that it involves higher derivatives.

In this paper we develop a method based on the Kostelecky-Berger (KB)
construction, to introduce a Lorentz-breaking deformation of the
supersymmetry (SUSY) algebra for supergauge field theories. Earlier,
this method was successfully applied to supersymmetric scalar
field theories \cite{Farias:2012ed}. There it was shown that the application of
the KB construction allows to generate aether-like terms \cite{aether},
in the action of the theories at the component level, while the
effective action can be calculated on the basis of the superfield
approach, in a way as simple as in the usual,
Lorentz-invariant case. We develop this methodology, both in three-
and in four-dimensional cases, and one of the key results of our
consideration consists in a natural arising of a new form of gauge
symmetry, involving the Lorentz-breaking parameter, for the vector
component of the  superfield.

The paper is organized as follows. In Sec. \ref{3daether} we discuss
the generalization of the three-dimensional aether superspace to gauge
theories, including some perturbative calculations on
three-dimensional supersymmetric quantum electrodynamics and
Chern-Simons-matter model. In Sec. \ref{4d-aether-superspace} we deal
with the four-dimensional case, and we apply the aether superspace
methodology in the computation of the effective potential to the
supersymmetric quantum electrodynamics. In Sec. \ref{coortra} we discuss the possibility of equivalence between our modification of the supersymmetry generators and some coordinate transformations. Finally, in Sec. \ref{summary} we
present our final  remarks.

\section{Three-dimensional aether superspace} \label{3daether}

Just as we have discussed in our previous paper~\cite{Farias:2012ed},
the extension of the usual superspace to a three-dimensional deformed
superspace is stated through the deformed SUSY  generators
\begin{eqnarray}
\label{modgen1}
Q_{\alpha}&=&i[\partial_{\alpha}-i\theta^{\beta}\gamma^m_{\beta\alpha}
(\partial_m+k_{mn}\partial^n)]\nonumber\\
&=&i[\partial_{\alpha}-i{\theta}^{{\beta}}\gamma^m_{\beta\alpha}\nabla_m],
\end{eqnarray}
\noindent satisfying the anti-commutation relation
\begin{eqnarray}
\{Q_{\alpha},Q_{\alpha}\}=2i\gamma^m_{{\alpha}\beta}\nabla_m,
\end{eqnarray}
\noindent where $\partial_{\alpha}$ is the derivative with respect to
the Grassmannian coordinates $\theta^{\alpha}$ and
$\nabla_m=\partial_m+k_{mn}\partial^n$, with $\partial_m$ the derivative with respect to $x^m$. 
Latin indices assume values of
three-dimensional space-time coordinates ($0,1,2$) and $k_{mn}$ is a
constant tensor which can be chosen to assume an aether-like form
$k_{mn}=\alpha u_mu_n$, with $\alpha$ a small
parameter (cf. \cite{aether}) and $u^m$ being a constant vector with
$u^mu_m$ equal either to 1, $-1$ or 0. 
In general, we use the conventions and
notations as well as normalization factors as in Ref.~\cite{SGRS}, but
the symbols for the (supergauge covariant) derivatives appearing in
the present paper are slightly  different.

It is important to remark that the new supercovariant derivative which 
anti-commute with $Q_\alpha$ is given by
\begin{eqnarray}
\label{sder3d}
D_{\alpha}=\partial_{\alpha}+i{\theta}^{{\beta}}\gamma^m_{{\beta}\alpha}\nabla_m~,
\end{eqnarray}
where the operator $\nabla_m$ commutes with $D_{\alpha}$, as well as
with the SUSY generators.

For superscalar field theories constructed in this deformed
superspace, we can define an  action such as
\begin{eqnarray}
\label{action3d}
S=-\frac{1}{2}\int{d^5z}\left[(\overline{D^{\alpha}\Phi})(D_\alpha\Phi)-
f(\bar{\Phi}\Phi)\right],
\end{eqnarray}
where $f(\bar{\Phi}\Phi)$ is some function of the bilinear
$\bar{\Phi}\Phi$, which is invariant under $U(1)$ global
transformations ($\Phi'=\mathrm{e}^{iK}\Phi$). Our aim in this paper
is to extend such action to theories that are invariant under local
(gauge) transformations ($\Phi'=\mathrm{e}^{iK(x,\theta)}\Phi$, with
$K(x,\theta)$ being a {\it{real}}  scalar superfield).

To do this, let us introduce a supergauge covariant derivative
$\mathcal{D}_{\alpha}=(D_\alpha-i\Gamma_\alpha)$ such tht
$\mathcal{D}_\alpha\Phi$ transforms covariantly under $U(1)$ gauge
transformations
($\mathcal{D}_{\alpha}'\Phi'=\mathrm{e}^{iK(x,\theta)}\mathcal{D}_{\alpha}\Phi$),
allowing us to write a gauge invariant action
\begin{eqnarray}\label{action3db}
S&=&-\frac{1}{2}\int d^5z\left[(\overline{\mathcal{D}^{\alpha}{\Phi}})
(\mathcal{D}_{\alpha}{\Phi})-f(\bar{\Phi}\Phi)\right]\nonumber\\
&=&-\frac{1}{2}\int
d^5z\left[(\overline{{D}^{\alpha}{\Phi}})({D}_{\alpha}{\Phi})-i
\overline{{D}^{\alpha}{\Phi}}\Gamma_\alpha\Phi+i\Gamma^\alpha\bar{\Phi}
{D}_{\alpha}{\Phi}
  +\Gamma^{\alpha}\Gamma_{\alpha}\bar{\Phi}\Phi- f(\bar{\Phi}\Phi)\right],
\end{eqnarray}
\noindent where the spinor gauge connection transforms as
$\Gamma_{\alpha}'=\Gamma_{\alpha}+D_\alpha K$. Note that the gauge
transformations themselves are defined as deformed  ones. 

The components of the spinor superfield connection can be defined as 
\begin{eqnarray}\label{compon1}
\chi_\alpha=\Gamma_\alpha\Big{|}_{\theta=0}~&,&~~B=\frac{1}{2}D^{\alpha}
\Gamma_\alpha\Big{|}_{\theta=0}~,\nonumber\\
V_{\alpha\beta}=-\frac{i}{2}D_{(\alpha}\Gamma_{\beta)}
\Big{|}_{\theta=0}~&,&~~\lambda_\alpha=\frac{1}{2}D^{\beta}D_{\alpha}\Gamma_\beta
\Big{|}_{\theta=0}~,
\end{eqnarray}

\noindent where $V_{\alpha\beta}=(\gamma^m)_{\alpha\beta}A_m$. In
analogy with the usual three-dimensional superspace, the components of
scalar superfield are conveniently  defined as
\begin{eqnarray}\label{compon2}
\varphi=\Phi\Big{|}_{\theta=0}~,~~\psi=D^{\alpha}\Phi
\Big{|}_{\theta=0}~,~~F=D^2\Phi\Big{|}_{\theta=0}~,
\end{eqnarray}

\noindent with similar definitions for the components of $\bar{\Phi}$.

Therefore, in terms of the components of the superfields, the action
Eq.(\ref{action3db})  can be cast as
\begin{eqnarray}
S&=&\int{d^3x}~\Big{\{}\bar{F}F+\bar{\psi}^\alpha{(\gamma^m)_{\alpha}}^{\beta}
[i\nabla_m-A_m]\psi_{\beta} 
+(i\bar{\psi}^{\alpha}\lambda_\alpha\varphi+h.c.)\nonumber\\
&&+(\nabla^m-iA^m)\bar{\varphi}(\nabla_m+iA_m)\varphi+\frac{1}{2}
f'(\bar{\varphi}\varphi)[\bar{F}\varphi+\bar{\varphi}F+2\bar{\psi}^\beta
\psi_\beta]\nonumber\\
&&+\frac{1}{2}f''(\bar{\varphi}\varphi)[2\bar{\varphi}\varphi\bar{\psi}^\beta
\psi_\beta+\varphi^2\bar{\psi}^\beta\bar{\psi}_\beta+\bar{\varphi}^2\psi^\beta
\psi_\beta]
\Big{\}}, \label{action3dbb}
\end{eqnarray}

\noindent where $f'(\bar{\varphi}\varphi)=
\dfrac{\partial f(\bar{\Phi}\Phi)}{\partial(\bar{\Phi}\Phi)}
\Big{|}_{\bar{\Phi}\Phi=\bar{\varphi}\varphi}$.

Gauge covariant superfield strength can be defined just as in the usual
case, $W_\alpha=\dfrac{1}{2}D^{\beta}D_{\alpha}\Gamma_{\beta}$. A SUSY
Maxwell Lorentz-breaking action can be constructed as
\begin{eqnarray}
S&=&\int d^5z\frac{1}{2}W^\alpha W_\alpha=\int{d^3x}
\left[\lambda^\alpha
  i{(\gamma^m)_{\alpha}}^{\beta}\nabla_m\lambda_\beta-\frac{1}{2}f^{\alpha\beta}
f_{\alpha\beta}\right],\label{action3dc} 
\end{eqnarray}

\noindent where $\lambda_\alpha\equiv W_\alpha|_{\theta=0}$ and
$f_{\alpha\beta}=D_\alpha W_\beta|_{\theta=0}=D_\beta
W_\alpha|_{\theta=0}$. In terms of the gauge field $A_m$,
$f_{\alpha\beta}$ can be written as 
$f_{\alpha\beta}=\dfrac{1}{2}~{\epsilon^{mn}}_{r}(\gamma^r)_{\alpha\beta}
\nabla_mA_n$. Therefore the physical content of the SUSY Maxwell-like action is given by
\begin{eqnarray}
S=\int{d^3x} \left[\lambda^\alpha i{(\gamma^m)_{\alpha}}^{\beta}\nabla_m\lambda_\beta-\frac{1}{4}(\nabla_mA_n-\nabla_nA_m)^2\right],\label{action3de}
\end{eqnarray}

\noindent where $\nabla_m=\partial_m+k_{mn}\partial^n$.

\subsection{Three-dimensional quantum electrodynamics in aether  
superspace}\label{aether-qed}

As a first example of the power of the superspace techniques even in
Lorentz breaking scenarios, let us evaluate the one-loop correction to
the self-energy of the gauge superfield in the Super Quantum
Electrodynamics in three-dimensions. To do this it is necessary to
compute the superpropagators of the model.  Considering the theory defined by the action 
Eq.(\ref{action3db}+\ref{action3dc}) plus the gauge-fixing and the 
corresponding Fadeev-Popov term, and proceeding as usual, we can write
the following propagators in the aether superspace
\begin{eqnarray}
\langle \Phi(p,\theta_1)\Phi(-p,\theta_2)\rangle&=&
\frac{(D^2-m)}{\tilde{p}^2+m^2}\delta^2(\theta_1-\theta_2)~,\nonumber\\
\langle \Gamma_\alpha(p,\theta_1)\Gamma_\beta(-p,\theta_2)\rangle&=&
\frac{1}{\tilde{p}^2}\left[\frac{(1+\xi)}{2}C_{\beta\alpha}
-\frac{(1-\xi)}{2}\frac{(\gamma^m)_{\beta\alpha}\tilde{p}_m~D^2}{\tilde{p}^2}
\right]\delta^2(\theta_1-\theta_2)~,\label{prop3d}
\end{eqnarray}  

\noindent where $\tilde{p}_m=p_m+k_{mn}p^n$,
$\tilde{p}^2=p^2+2k_{mn}p^mp^n+k^{mn}k_{ml}p_np^l$,
$D^2=\partial^2-\theta^\beta(\gamma^m)_{\beta\alpha}\tilde{p}_m\partial^\alpha
+\theta^2\tilde{p}^2$. As we commented in our previous
work~\cite{Farias:2012ed}, this dispersion relation has a structure
common for the propagators in the CPT-even Lorentz-breaking theories 
(see  e.g.~\cite{Fer}).

\begin{figure}[h]
\includegraphics[]{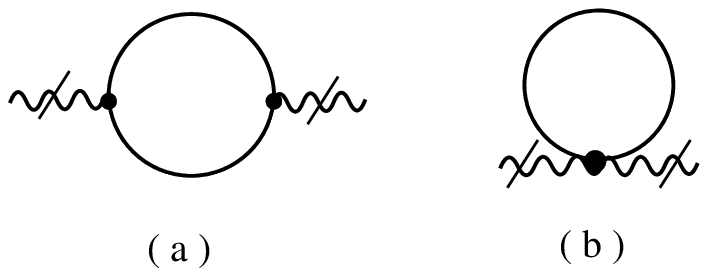}
\caption{One-loop contributions to the gauge superfield effective action. Continuous lines represent the scalar superfield propagator, and wave crossed lines represent the external gauge superfield.}\label{fig1}
\end{figure}

We are able to compute the radiative corrections to the Super Quantum
Electrodynamics Lorentz-breaking theory, Eq.(\ref{action3dc}). and
choosing $f(\bar{\Phi}\Phi)=M\bar{\Phi}\Phi$ (i.e., a mass term to the
scalar superfield), the diagrams which contribute to the effective
action is depicted in Fig.\ref{fig1}. The corresponding expression can
be  cast as   
\begin{eqnarray}
S_{2l}&=&\int{\frac{d^3p}{(2\pi)^3}d^2\theta}\left[W^\alpha
  W_\alpha-M~\Gamma^\alpha W_\alpha\right]\int{\frac{d^3q}{(2\pi)^3}}
\frac{1}{(\tilde{q}^2+M^2)[(\tilde{q}-\tilde{p})^2+M^2]}.\label{eff-action1}
\end{eqnarray}

\noindent This last integral can be evaluated by changing the variable of
integration $q$ to $\tilde{q}$. In the case of $p^2\approx0$ we can write
$\int{d^3q}=\Delta\int{d^3\tilde{q}}$, where
$\Delta=\det{(\frac{\partial
    q^m}{\partial\tilde{q}^n})}=\det^{-1}(\delta^m_n+k^m_n)$ is the
Jacobian of the changing of variables. For $k_{mn}=\alpha
u_mu_n$ with a small $\alpha$,
$\Delta\approx(1-\alpha u^2)$. So, the final result is 
\begin{eqnarray}
\label{fin3d}
S_{2l}&=&\frac{\Delta}{8\pi|M|}\int{\frac{d^3p}{(2\pi)^3}d^2\theta}\left[W^\alpha
  W_\alpha-
M~\Gamma^\alpha W_\alpha\right].\label{eff-action1a}
\end{eqnarray}

We can observe that the one-loop quantum correction is finite. This
model is known to be finite to all loop orders in perturbation theory
in the usual superspace~\cite{Piguet,Ferrari:2007mh}, and it is
natural to expect that this issue persists in the aether superspace,
since the power counting of the model is not affected by the presence
of the Lorentz breaking terms introduced through the aether
superspace. We also observe, the generation of a Super-Chern-Simons
Lorentz-breaking term, which corresponding bosonic local part has the
form
\begin{eqnarray}
\int{d^3x}~M~\Delta~\epsilon^{lmn}A_l\nabla_m A_n=\int{d^3x}~M~
\Delta~\epsilon^{lmn}[A_l\partial_mA_n+k_{ms}A_l\partial^s A_n].\label{csterm}
\end{eqnarray}
We note that the Chern-Simons action, instead of the usual gauge 
transformations, is invariant under the new ones $\delta A_n=\nabla_n\xi$, 
with $\xi$ being a parameter of the gauge transformation.

Let us now, discuss the Maxwell action. It is easy to see that, after doing the Fourier transformation and reducing to the component fields, the Maxwell-like contribution from (\ref{fin3d}) looks like
\begin{eqnarray}
\label{accomp}
S_M=-\frac{1}{4}\frac{\Delta}{8\pi|M|}\int d^3x\tilde{F}_{mn}\tilde{F}^{mn},
\end{eqnarray}
where 
\begin{eqnarray}
\label{newtens}
\tilde{F}_{mn}=\nabla_m A_n-\nabla_n A_m
\end{eqnarray}
is a new stress tensor. We note that the derivatives $\nabla_m$ emerge from the supercovariant spinor derivatives.
So, as can be seen, a new action for the vector field
$A_m$ is generated, that is (as the Chern-Simons one) invariant under the new
gauge transformations $\delta A_n=\nabla_n\xi$. We note that the action
(\ref{accomp}) essentially differs from the usual aether-like action
for the gauge field \cite{aether} which is invariant under the usual
gauge transformations  $\delta A_n=\partial_n\xi$, and cannot be reduced to it.

\subsection{Chern-Simons-matter model in aether Superspace}
\label{aether-csm}

Quantum field theories defined in a three-dimensional space-time are
widely discussed in the literature because they offer a very rich
structure, working as excellent theoretical laboratories as well as
they can be applied to some almost planar condensed matter systems,
such as quantum Hall effect~\cite{prange}. More recently,
supersymmetric gauge field theories in three dimensions could be
related to
M2-branes~\cite{Bagger:2006sk,Bagger:2007jr,Gustavsson:2007vu}. Of
special interest is the computation of the effective superpotential of
the Supersymmetric Chern-Simons-matter
model~\cite{Gomes:2012qv,Lehum:2012mp,Lehum:2010tt,Ferrari:2010ex,
Buchbinder:2010em,
  Buchbinder:2009dc}, that can be used to evaluate the possibility of
spontaneous (super)symmetry breaking via Coleman-Weinberg
mechanism~\cite{Coleman:1973jx}. The presence of
Lorentz symmetry violating terms in the Lagrangian could be a source
of spontaneous SUSY  breaking~\cite{Katz:2006rx} .

Let us start by defining the classical action of the model
\begin{eqnarray}
S&=&\int{d^5z}\Big{\{}\Gamma^{\alpha}W_{\alpha}
-\frac{1}{2}\overline{\mathcal{D}^{\alpha}\Phi}\mathcal{D}_{\alpha}\Phi
+
\lambda(\bar\Phi\Phi)^2\Big{\}},\label{ceq1}
\end{eqnarray}

\noindent
where  $W^{\alpha}=(1/2)D^{\beta}D^{\alpha}\Gamma_{\beta}$ is the
gauge superfield strength as defined before and
$\mathcal{D}^{\alpha}=(D^{\alpha}-ie\Gamma^{\alpha})$ is the
supercovariant  derivative.

The action Eq.(\ref{ceq1}) possesses manifest ${\cal{N}} = 1$ SUSY,
and it can be lifted to ${\cal{N}}=2$ by the elimination of the
fermion-number violating terms~\cite{Lee:1990it}, from which we
identify the coupling constants as $\lambda=-e^2/8$. In the usual
superspace, SCSM is superconformal invariant at classical level, but
in the aether superspace the presence of the constant vector $k_{mn}$
explicitly breaks this invariance; even so, we should expect that an analog
(or extended) symmetry could emerge from the action
Eq.(\ref{ceq1}), but we will not extend such analysis in this  paper.

The presence of a mass term like $\int{d^5z}M\bar\Phi\Phi$, with a wrong sign, in the
action Eq.(\ref{ceq1}) would generate a
spontaneously (gauge) symmetry broken phase at classical
level~\cite{Lehum:2007nf}, but here we are interested in spontaneous
symmetry breaking induced by radiative corrections (Coleman-Weinberg
mechanism), and so we will keep the model massless at this level. To
this end, let us shift the superfields $\bar\Phi$ and $\Phi$ by the
classical background superfield $\varphi$  as
\begin{eqnarray}
\bar\Phi&=&\frac{1}{\sqrt{2}}\Big(\varphi+\Phi_1-i\Phi_2\Big)\nonumber\\
\Phi&=&\frac{1}{\sqrt{2}}\Big(\varphi+\Phi_1+i\Phi_2\Big)~,\label{ceq5}
\end{eqnarray}

\noindent where $\varphi=\varphi_1-\theta^2\varphi_2$, with $\varphi_1$
and $\varphi_2$ being  real constant fields. 

Assuming the vanishing of the vacuum expectation values (VEV) of the
quantum superfields, i.e.,
$\langle\Phi_1\rangle=\langle\Phi_2\rangle=0$ at any order of
perturbation theory, the gauge
invariant action Eq.(\ref{ceq1}) results  in,
\begin{eqnarray}
S&=&\int{d^5z}\Big{\{}\Gamma^{\alpha}W_{\alpha}-\frac{e^2\varphi^2}{4}
\Gamma^{\alpha}\Gamma_{\alpha}
-\frac{e\varphi}{2}D^{\alpha}\Gamma_{\alpha}\Phi_2
+\frac{1}{2}\Phi_1(D^2+3\lambda\varphi^2)\Phi_1+\frac{1}{2}\Phi_2(D^2+
\lambda\varphi^2)\Phi_2\nonumber\\
&+&\frac{1}{2}\varphi D^2\varphi+\frac{\lambda}{4}\varphi^4
+\frac{e}{2}D^{\alpha}\Phi_2 \Gamma_{\alpha}\Phi_1
-\frac{e}{2}D^{\alpha}\Phi_1 \Gamma_{\alpha}\Phi_2
-\frac{e^2}{2}(\Phi_1^2+\Phi_2^2)\Gamma^2-e^2\varphi\Phi_1 \Gamma^2\nonumber\\
&+&\frac{\lambda}{4}(\Phi_1^4+\Phi_2^4)
+\frac{\lambda}{2}\Phi_1^2\Phi_2^2+\lambda
\varphi\Phi_1(\Phi_1^2+\Phi_2^2)-
eD^{\alpha}\varphi \Phi_2 \Gamma_{\alpha}
+(\lambda \varphi^3+D^2\varphi)\Phi_1 \nonumber\\
&+& \frac{1}{2\alpha} (D^{\alpha}\Gamma_{\alpha}+\alpha\frac{e\varphi}{2}\Phi_2)^2
+\bar{c}D^2c+\frac{\alpha}{4}e^2\varphi^2\bar{c}c+\frac{\alpha}{4}{e^2\varphi}
\bar{c}\Phi_1 c
\Big{\}}.\label{ceq6}
\end{eqnarray}

\noindent In the last line we added a gauge fixing and the
corresponding Faddeev-Popov terms. We have used an $R_{\xi}$ gauge
condition to eliminate the mixing between $\Gamma$ and $\Phi_2$
superfields, but this procedure is not enough to completely eliminate
this mixing. Even so, the remaining term, $-e D^{\alpha} \varphi
\Phi_2 \Gamma_{\alpha}$, can be disregarded in the K\"{a}hlerian
approximation of the effective superpotential, because it contains a
supercovariant derivative applied to the background superfield $\varphi$. 

The knowledge of the K\"{a}hlerian effective superpotential is enough
to determine the possibility of spontaneous SUSY and gauge symmetry
breaking~\cite{Alv,Ferrari:2010ex}. 
We will evaluate it at two-loop order, where such effects are expected
to show up~\cite{Ferrari:2010ex,Dias:2003pw,Tan:1996kz}.    

The Feynman rules derived from Eq.(\ref{ceq6}) are given, in the K\"{a}hlerian
approximation (that is, by preserving the dependence in $\varphi$ and
dropping the dependences on $D_{\alpha}\varphi$ and $D^2\varphi$), by 
\begin{eqnarray}\label{props}
\langle T~\Phi_1(k,\theta)\Phi_1(-k,\theta')\rangle&=&-i
\frac{D^2-M_{\Phi_1}}{k^2+M_{\Phi_1}^2}\delta^{(2)}(\theta-\theta')~,\nonumber\\
\langle T~\Phi_2(k,\theta)\Phi_2(-k,\theta')\rangle&=&-i
\frac{D^2-M_{\Phi_2}}{k^2+M_{\Phi_2}^2}\delta^{(2)}(\theta-\theta')~,\\
\langle
T~\Gamma_{\alpha}(k,\theta)\Gamma_{\beta}(-k,\theta')\rangle&=&
\frac{i}{4}
\Big[\frac{(D^2+M_{A})D^2D_{\beta}D_{\alpha}}{ k^2(k^2+M_{\Gamma}^2)}\nonumber\\
&+&\alpha\frac{(D^2-\alpha M_{\Gamma})D^2D_{\alpha}D_{\beta}}
{k^2(k^2+\alpha^2M_{\Gamma}^2)}\Big]\delta^{(2)}(\theta-\theta')~.\nonumber
\end{eqnarray}

For simplicity, let us choose the SUSY Landau gauge $\alpha=0$ (we
have to remark that the effective superpotential is a gauge-dependent
quantity~\cite{Jackiw:1974cv}). With this choice, the ghost
superfields are decoupled from the model, and we can identify the
poles of the propagators of the interacting superfields  as
\begin{eqnarray}\label{mass}
M_{\Phi_1}=3\lambda\varphi^2, \quad M_{\Phi_2}=\lambda\varphi^2~,
\quad 
M_{\Gamma}=\frac{e^2\varphi^2}{4}~.
\end{eqnarray}

\begin{figure}[h] 
\includegraphics[angle=90,width=7cm]{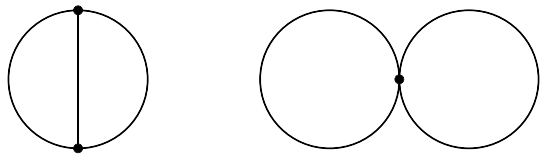} \caption{Topologies of two-loop diagrams that contribute to the K\"{a}hlerian effective superpotential.}\label{fig2} \end{figure}

Proceeding as described in \cite{Ferrari:2009zx}, considering the
two-loop corrections depicted in Fig \ref{fig2}, and performing the
integrals using the regularization by dimensional reduction
~\cite{Siegel:1979wq}, the two-loop K\"{a}hlerian effective
superpotential can be cast  as
\begin{eqnarray}\label{ceq11a}
K(\varphi)&=&-\frac{b_2}{4}\varphi^4\left(\frac{b_1}{b_2}-\frac{1}{2}~\Delta^2+
\Delta^2~\ln{\frac{\varphi^2}{\mu}}\right)-\frac{B}{4}\varphi^4~.
\end{eqnarray}

\noindent where $B$ is a counterterm, $\mu$ is a mass scale introduced
by the regularization, $b_1$ is a function of the coupling constants,
of $~\Delta$ and of $1/\epsilon\equiv \frac{1}{3-D}$ ($D$ is the
dimension of 
the space-time). The quantity $b_2$ is explicitly given by 
\begin{eqnarray}\label{ceq8a}
b_2 &=& -(116~e^6+543~e^4\lambda+432e^2\lambda^2-
71552~\lambda^3)/(12288\pi^2).
\end{eqnarray}

The counterterm $B$ is fixed through the following  renormalization condition
\begin{eqnarray}
\label{cweq9}
\frac{\lambda}{4}\equiv\frac{1}{4!}\frac{\partial^4
  K(\varphi)}{\partial 
\varphi^4}\Big{|}_{\varphi=v}~,
\end{eqnarray}

\noindent
where $v$ is the renormalization point. By substituting $B$ in Eq.
(\ref{ceq11a}), the K\"{a}hlerian effective superpotential results in
\begin{eqnarray}\label{ceq11}
K(\varphi)=-\frac{b_2\Delta^2}{4}\varphi^4
\ln\left[\frac{\varphi^2}{v^2}\exp\left(-\frac{\lambda}{b_2\Delta^2}-\frac{25}{6}
\right)\right]~.
\end{eqnarray}

Now we are able to study the spontaneous generation of mass to the
physical superfields induced by the radiative corrections. First, let
us impose the condition to  extremize the K\"{a}hlerian effective 
superpotential. It reads
\begin{eqnarray}\label{ceq11aa}
\frac{\partial K(\varphi)}{\partial\varphi}=
\dfrac{\varphi^3}{3}\left[
3\lambda+3b_2\Delta^2\left(\dfrac{11}{3}-\ln\dfrac{\varphi^2}{v^2}\right) 
\right]=0.
\end{eqnarray}

\noindent The nontrivial solutions are given by
\begin{eqnarray}\label{ceq11ab}
\varphi=\pm v~\exp\left(\dfrac{11}{6}+\dfrac{\lambda}{2 b_2\Delta^2}\right).
\end{eqnarray}

Since we have computed the effective superpotential for constant
configurations of the background superfields, we expect that our
approximation is valid for small fluctuations around the mass scale
$v$, used as renormalization point. This expectative constraints the
exponential of the above equation to be approximately $1$. Therefore,
we want that the coupling constants satisfy 
$\dfrac{11}{6}+\dfrac{\lambda}{2 b_2\Delta^2}\approx 0$, what results in
the following condition
\begin{eqnarray}\label{ceq11ac}
\lambda\approx \dfrac{11}{3}b_2\Delta^2 \approx
\left[-(4\times10^{-3})~e^6 -(16\times 10^{-3})~e^4\lambda -(13\times
  10^{-3})~e^2\lambda^2 +2\lambda^3\right]
\Delta^2.
\end{eqnarray}

For the Coleman-Weinberg mechanism, this last equation is directly
related to the compatibility of the effective superpotential
calculations with the assumptions of perturbation theory. We can see
from Eq.~(\ref{ceq11ac}) that $\lambda$ should be of order of
$(4\times10^{-3})e^6\Delta^2+\mathcal{O}(e^{10})$, so that, for
small $e$ we are in the regime of validity of the perturbative expansion.

The second  derivative of the K\"{a}hlerian effective superpotential
with respect to the background field $\varphi$ evaluated in the minimum
of the superpotential, i.e. $\varphi\simeq \pm v$, is interpreted as
the mass of the matter (background) superfield $\varphi$. If positive,
this condition guarantees that Eq.(\ref{ceq11ab}) is a minimum of
$K(\varphi)$. In fact, using Eqs. (\ref{ceq11ac}) and (\ref{ceq11ac}),
we obtain
\begin{eqnarray}\label{ceq11ad}
M_{\varphi}&=&\frac{d^2K(\varphi)}{d\varphi^2}\Big|_{\varphi=v}
\approx{(2\times 10^{-3})e^6}v^2\Delta^2~,
\end{eqnarray}

\noindent
and the mass of the gauge superfield induced by the radiative
corrections is 
given by 
\begin{eqnarray}\label{ceq12}
\dfrac{M_\Gamma}{M_{\varphi}}\sim\frac{e^2}{12\lambda}~\sim~ 
-21\dfrac{e^{-4}}{\Delta^2}~,
\end{eqnarray}

\noindent where we can notice that the mass of the gauge superfield is
much larger than the mass of matter superfield since, for a small
violation of the Lorentz symmetry, $\Delta$ should be approximately $1$.

One interesting remark is that all information of the presence of
Lorentz violating terms in the original action is manifested in the
presence of the $\Delta$ factor in the effective superpotential, and
consequently in the induced masses. Since the K\"{a}hlerian effective
superpotential has a consistent minimum, we can affirm that SUSY can
not be spontaneously broken via Coleman-Weinberg mechanism. Probably,
to search for SUSY breaking induced by the Lorentz violating terms we
should compute the whole effective action, using some more
sophisticated  technology.  

\section{Four-dimensional aether superspace}
\label{4d-aether-superspace}

Now, let us consider gauge theories within the four-dimensional
aether  superspace. In this case, the spinor supercovariant
derivatives look  like
\begin{eqnarray}
\label{sder}
D_{\alpha}=\partial_{\alpha}+i\bar{\theta}^{\dot{\beta}}\sigma^m_{\dot{\beta}\alpha}
\nabla_m;\nonumber\\
\bar{D}_{\dot{\alpha}}=\partial_{\dot{\alpha}}+i\theta^{\beta}
\bar{\sigma}^m_{\beta\dot{\alpha}}\nabla_m,
\end{eqnarray}
where again $\nabla_m=\partial_m+k_{mn}\partial^n$.
It is clear that these spinor derivatives satisfy the usual properties
\begin{eqnarray}
D_{\alpha}D_{\beta}D_{\gamma}=0;\quad\,
\bar{D}_{\dot{\alpha}}\bar{D}_{\dot{\beta}}
\bar{D}_{\dot{\gamma}}=0.
\end{eqnarray}
Now, let us define the following Abelian gauge theory
\begin{eqnarray}
S=\int d^6z W^{\alpha} W_{\alpha},\label{1}
\end{eqnarray}
where
\begin{eqnarray}
W_{\alpha}=\frac{1}{8}\bar{D}^2(e^{-v}D_{\alpha}e^v)=
\frac{1}{8}\bar{D}^2
D_{\alpha}v.
\end{eqnarray}
In principle, the non-Abelian generalization of this theory can be
constructed along the same lines.

This action can be rewritten as
\begin{eqnarray}
\label{sw}
S_W=-\frac{1}{16}\int d^8z v D^{\alpha}\bar{D}^2D_{\alpha} v.
\end{eqnarray}
As can be seen, its form does not differ from the usual action of
gauge theories (see f.e. \cite{SGRS}); the only difference, from the
usual case, consists in the replacement of the common spinor
supercovariant derivative by a new one given by (\ref{sder}). It is
clear that this action is invariant under the gauge transformations
$\delta v=\Lambda+\bar{\Lambda}$, where $\Lambda$ is a chiral
superfield, and $\bar{\Lambda}$ is an antichiral  one.

Following the general principles, we suggest that the component
expansion of the real scalar superfield $v$ is the same as in the
usual case, i.e. it depends on the relevant vector (gauge) field $A_m$
as
\begin{eqnarray}
v=-\frac{i}{2}(\bar{\theta}\sigma^m\theta)A_m(x)+\ldots.
\end{eqnarray}
By reducing the action (\ref{sw}) to the component fields, its
bosonic part can be shown to have the form (\ref{accomp})  up to the numerical factor, with the only difference that the integral is now performed over the four-dimensional space-time.

Then, we must add the following gauge-fixing action
\begin{eqnarray}
S_{gf}=\frac{1}{16\alpha}\int d^8z v D^2\bar{D}^2v,
\end{eqnarray}
where $\alpha$ is the gauge-fixing parameter.

The corresponding propagator looks like
\begin{eqnarray}
\label{propv}
&&<v(z_1)v(z_2)>=-\frac{1}{\tilde{\Box}}
(-\frac{D^{\alpha}\bar{D}^2D_{\alpha}}{8\tilde{\Box}}+
\alpha\frac{\{\bar{D}^2,D^2\}}{16
\tilde{\Box}})
\delta^8(z_1-z_2),
\end{eqnarray}
involving the new projection operators 
$$
\Pi_0=\frac{\{\bar{D}^2,D^2\}}{16\tilde{\Box}},\quad\,\Pi_{1/2}=-
\frac{D^{\alpha}
  \bar{D}^2D_{\alpha}}{8\tilde{\Box}}.
$$
Then, we couple the gauge field to the chiral matter filed $\phi$, by introducing
the following  action
\begin{eqnarray}
S_{\Phi}=\int d^8z \bar{\phi}e^{gv}\phi.
\end{eqnarray}
The propagators of the chiral field look like (cf. \cite{Farias:2012ed}):
\begin{eqnarray}
\label{props1}
&&<\phi(z_1)\bar{\phi}(z_2)>=\frac{\bar{D}^2D^2}{16\tilde{\Box}}
\delta^8(z_1-z_2)\,\,\,\,\,\,\,
<\bar{\phi}(z_1)\phi(z_2)>=\frac{D^2\bar{D}^2}{16\tilde{\Box}}\delta^8(z_1-z_2).
\end{eqnarray}
To calculate the one-loop K\"{a}hlerian effective potential, we can
use the well-developed methodology of calculating the superfield
effective potential elaborated in  \cite{YMEP,efpot,BCP}. 
As usual, one can begin with constructing the one-loop Feynman
diagrams contributing to the superfield effective potential. The
structure of the supergraphs does not essentially differ from the
usual case  \cite{YMEP}.
The first set depicted at Fig. 3 involves only gauge
propagators. Their sum is  given by
\begin{eqnarray}
K^{(1)}_a=\int
d^8z_1\sum\limits_{n=1}^{\infty}\frac{(-1)^n}{2n}(g^2\Phi\bar{\Phi}
\frac{1}{\Box}(\Pi_{1/2}+\alpha\Pi_0))^n\delta_{12}|_
{\theta_1=\theta_2},
\end{eqnarray}
where $\frac{1}{n}$ is a symmetry factor. 

Proceeding just as in \cite{YMEP}, we find that
\begin{eqnarray}
\label{k1a}
K^{(1)}_a=\int
d^8z\int\frac{d^4p}{(2\pi)^4}\frac{1}{\tilde{p}^2}\Big[\ln(1+
\frac{g^2\Phi\bar{\Phi}
}{\tilde{p}^2})-\ln(1+\frac{\alpha
  g^2\Phi\bar{\Phi}}{\tilde{p}^2}) \Big],
\end{eqnarray}
where $\tilde{p}^2=(p_m+k_{mn}p^n)^2$ is a Fourier transform for $\tilde{\Box}$.
Notice that at $\alpha=0$ (Landau gauge), the second term in
(\ref{k1a})  
vanishes. Using the
notations adopted in \cite{hd}, one can introduce a ``dressed''
propagator involving a sum over quartic  vertices (see Fig. 4):
\begin{eqnarray}
<vv>_D=-(\frac{1}{\tilde{\Box}+g^2\Phi\bar{\Phi}}\Pi_{1/2}+
\frac{\alpha}{\tilde{\Box}+\alpha g^2\Phi\bar{\Phi}} \Pi_0)\delta^8(z_1-z_2).
\end{eqnarray}

The triple vertices will enter the Feynman diagrams only through the
links depicted at Fig. 5, and the contribution from this sector is
given by the sum of the supergraphs depicted at Fig. 6. It is  equal to
\begin{eqnarray}
K^{(1)}_b=\int d^8z\int\frac{d^4p}{(2\pi)^4}\frac{1}{\tilde{p}^2}\Big[
\ln(1+\frac{\alpha g^2\Phi\bar{\Phi}}{\tilde{p}^2})\Big].
\end{eqnarray}
The total result which is the sum of $K^{(1)}_a$ and $K^{(1)}_b$, is gauge
invariant  and equal to
\begin{eqnarray}
\label{k1t}
K^{(1)}=\int
d^8z\int\frac{d^4p}{(2\pi)^4}\frac{1}{\tilde{p}^2}\ln(1+\frac{g^2\Phi\bar{\Phi}
}{\tilde{p}^2}).
\end{eqnarray}
To calculate these integrals, we can change the variables as in
\cite{Farias:2012ed}. After integration and subtracting the divergences we arrive at
\begin{eqnarray}
K^{(1)}=-\frac{1}{32\pi^2}\Delta g^2\Phi\bar{\Phi}
\ln\frac{g^2\Phi\bar{\Phi}}{\mu^2},
\end{eqnarray}
where $\Delta$ is again a Jacobian of the change of variables $k^m \rightarrow \tilde k^m$. We see that
the result only differs from the usual case \cite{efpot}, by the multiplicative factor $\Delta$.

\section{Lorentz-breaking modification of the supersymmetry generators and coordinate transformations}\label{coortra}

 To close the paper, let us discuss the possible impacts of the Lorentz-breaking modification of the supersymmetry generators for the generic quantum contributions to an effective action of arbitrary superfield theory. It follows from the the definitions of modified supersymmetry generators and covariant derivatives (\ref{modgen1},\ref{sder}) that the methodology of Lorentz symmetry breaking adopted by us implies
 the change of all momenta, that emerge through the D-algebra transformations, by the rule $p_m\to p_m+k_{mn}p^n$, or, in the coordinate space, $\partial_m\to \partial_m+k_{mn}\partial^n$ (which corresponds to the linear coordinate change $x_m\to (\delta_m^n+k_m^n)^{-1}x_n$). Therefore one can elaborate the following geometric interpretation of this Lorentz-breaking modification of SUSY algebra.

It was shown in our previous paper \cite{Farias:2012ed} that the one-loop contribution to two-point function in a $3D$ self-coupled scalar superfield model looks like
\begin{eqnarray}
\Gamma^{(1)}_2=\frac{\Delta}{8\pi|m|}\int d^3xd^2\theta \Phi(D^2-2m)\Phi.
\end{eqnarray}
By projecting this action to components, we arrive at
\begin{eqnarray}
\label{compcor}
\Gamma^{(1)}_2=\frac{\Delta}{8\pi|m|}\int d^3x (-\eta^{mn}\nabla_m\phi\nabla_n\phi+\psi^{\alpha}i(\gamma^m)_{\alpha}^{\beta}\nabla_m\psi_{\beta}+F^2-2m(\psi^2+\phi F)).
\end{eqnarray}
Let us perform, for this action, the analysis carried out in \cite{CollMac}
It is clear that one can formally introduce the upper-index metric
\begin{eqnarray}
\label{metric}
g^{ab}=\eta^{mn}(\delta_m^a+k_m^a)(\delta_n^b+k_n^b),
\end{eqnarray}
with $g_{ab}$ introduced as usual to be the inverse of $g^{ab}$. The Jacobian $\Delta=det^{-1}(\delta_a^m+k_a^m)$ can naturally be treated as a contribution to the integral measure, since $\Delta=\sqrt{|det\, g_{ab}|}=\sqrt{|g|}$. One can also introduce the modified Dirac matrices $\tilde{\gamma}^m=\gamma^m+k^m_n\gamma^n$ which satisfy the modified anticommutation relation
\begin{eqnarray}
\{\tilde{\gamma}^a,\tilde{\gamma}^b\}=2g^{ab},
\end{eqnarray}
where $g^{ab}$ is given by (\ref{metric}). Therefore, the action (\ref{compcor}) can be rewritten as
\begin{eqnarray}
\label{compcor1}
\Gamma^{(1)}_2=\frac{1}{8\pi|m|}\int d^3x\sqrt{|g|} (-g^{mn}\partial_m\phi\partial_n\phi+\psi^{\alpha}i(\tilde{\gamma}^m)_{\alpha}^{\beta}\partial_m\psi_{\beta}+F^2-2m(\psi^2+\phi F)).
\end{eqnarray}
Therefore, we can say that, for the scalar superfield, our Lorentz-breaking modification of the supersymmetry generators is equivalent to introduction of a new metric, and, therefore, of a new geometry (this is an affine geometry since the new metric is related to the Minkowski one through a constant matrix). It is easy to check that an analogous situation occurs also in the four-dimensional chiral superfield theory, that is, in our extension of the Wess-Zumino model.

However, the situation differs for the contributions involving external gauge legs. We have shown above that in the $3D$ gauge theory, the quantum correction is given by the expression (\ref{fin3d}).
If we project it into components, the result in the purely gauge sector will be
\begin{eqnarray}
\label{accomp1}
S_M=\frac{\Delta}{8\pi|M|}\int d^3x(-\frac{1}{4}\eta^{ma}\eta^{nb}\tilde{F}_{mn}\tilde{F}_{ab}-M\epsilon^{abc}A_a\nabla_b A_c).
\end{eqnarray}
Let us consider the Maxwell term, and, more precisely, one contribution to it, for example,
\begin{eqnarray}
\label{accomp2}
S_{M1}=\frac{\Delta}{8\pi|M|}\int d^3x(-\frac{1}{4})\eta^{ma}\eta^{nb}\nabla_mA_n\nabla_aA_b.
\end{eqnarray}
Repeating identically the arguments above, we can rewrite this expression as
\begin{eqnarray}
\label{accomp3}
S_{M1}=\frac{1}{8\pi|M|}\int d^3x\sqrt{|g|}(-\frac{1}{4})g^{ma}\eta^{nb}\partial_mA_n\partial_aA_b.
\end{eqnarray}
We see that while we succeeded in to replace the Minkowski metric to a new metric $g^{ab}$ in a sector involving only the space-time derivatives, there is no manner of to form a new "curved" metric $g^{ab}$ in a sector involving vector fields. This is related to the fact that within our methodology only the geometry (that is, coordinates, derivatives, metric and Dirac matrices) suffer transformations due to the introduction of the Lorentz-breaking parameters $k_{ab}$, but not the vector fields.
 Thus, there is no way to re-absorb the Lorentz breaking completely within a corresponding coordinate transformation in the gauge sector. This conclusion is similar to that one performed in \cite{CollMac}.

\section{Summary}\label{summary}

In this work we developed a gauge superfield method to construct
Lorentz-breaking supersymmetric field theories based on the
Kostelecky-Berger construction \cite{KostBer}. The methodology of
superfields is a powerful tool to study, among others, perturbative
aspects of supersymmetric theories. Even though SUSY and its algebra
is closely related to Lorentz symmetry, we could extend the superfield
formalism to include the Lorentz violating terms, therefore
allowing us to use the most attractive properties of superspace
formalism. The aether superspace is a natural way to deal with Lorentz
violating supersymmetric models. In this context, we presented some
applications of the aether superspace techniques in three and
four-dimensional space-time, discussing perturbative aspects of
supersymmetric quantum electrodynamics and Super-Chern-Simons-matter
model. We showed that, from the methodological viewpoint, the
calculations do not essentially differ from the usual
Lorentz-invariant  case. However, as we have noted, the new theory involving the Lorentz symmetry breaking can be reduced, though simple rules, to the usual Lorentz invariant theory, only if it is being considered in the purely scalar sector. If one deal with vector or spinor fields whose action involve metric contracted to fields, the redefinition of coordinates will not allow to redefine completely all the action, since it will imply in variations of the fields which are not suggested by the initial structure of our modification of the supersymmetry generators, cf. \cite{CollMac}.

\vspace{1cm}

{\bf Acknowledgments.}
This work was partially supported by Conselho Nacional de
Desenvolvimento Cient\'{\i}fico e Tecnol\'{o}gico (CNPq),
Funda\c{c}\~ao de Amparo \`a Pesquisa do Estado de S\~ao Paulo
(FAPESP) and Funda\c{c}\~{a}o de Apoio \`{a} Pesquisa do Rio Grande do
Norte (FAPERN). A. Yu. P. has been supported by the CNPq project  303438/2012-6.

\begin{figure}[ht]
\includegraphics[]{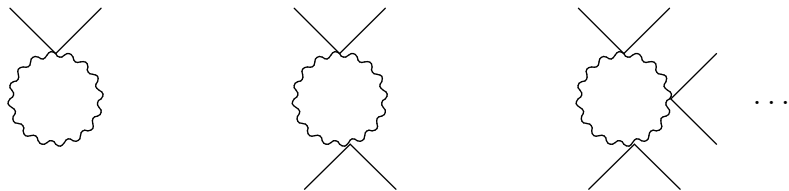}
\caption{Supergraphs composed by gauge propagators only.}\label{fig3}
\end{figure}

\begin{figure}[ht]
\includegraphics[]{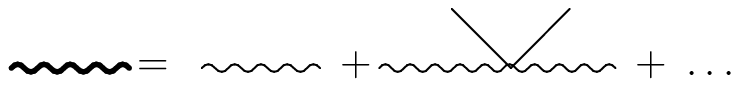}
\caption{Dressed propagator.}\label{fig4}
\end{figure}

\begin{figure}[ht]
\includegraphics[]{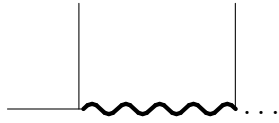}
\caption{A link involving gauge and matter propagators.}\label{fig5}
\end{figure}

\begin{figure}[ht]
\includegraphics[]{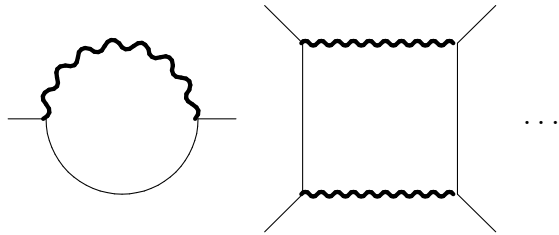}
\caption{Supergraphs composed by gauge and matter propagators.}\label{fig6}
\end{figure}

\end{document}